# New Technologies, Training Initiatives and the Future of Manuscript Studies

Eyal Poleg, Queen Mary University of London (e.poleg@qmul.ac.uk)

*We are standing at the edge of a major transformation in manuscript studies. Digital surrogates, Digital Humanities analyses and the rise of new scientific analytical technologies proliferate across universities, libraries and museums. They change the way we consult, research and disseminate historical manuscripts to reveal hitherto unknown, and unknowable, information. This article looks at how the field can best integrate these transformations. Concentrating on training programmes for advanced students as a way of reimagining the field, it provides concrete advice for the future of manuscript studies, arguing that the existence of manuscript studies as removed from Digital Humanities and heritage science is becoming more and more artificial and detrimental to the future of the field.*

The future of manuscript studies is already here. Digital images of manuscripts are proliferating online, becoming a bedrock of research and teaching, as well as inspiring creative memes and social media engagement. New technologies irreversibly change the way we consult, study, and communicate about manuscripts – ancient, medieval and modern alike. Covid outbreaks have made this even more pronounced, with the need to conduct teaching and research remotely. The rise of non-invasive analytical technologies, big-data tools and the availability of enhanced computing and imaging facilities, combine to transform the technological study of manuscripts. The possibilities embedded in digital and technological tools for studying, communicating, and teaching manuscripts have been put forward in recent conferences such as AMARC 2023 Spring Meeting on Heritage Science or the conference *On the Way to the Future of Digital Manuscript Studies* (Nijmegen, 2021), which served as the impetus for the current work.[1] The field, however, seems to be adjusting quite reluctantly to this new reality and in many universities the training of the next generation of scholars seems to put aside many of the major transformations of the past decades.

At its core manuscript studies is a complex and interdisciplinary field. Students are inducted to become inter- and multi-disciplinary scholars, in a field that builds upon a variety of linguistic skills, as well as auxiliary sciences necessary for analyzing the manuscript's creation and use. Palaeography and codicology explore the script and the physical construction of the book, while philology and linguistic analysis address the textual data encoded in its leaves. In tandem with the rise of book history, scholars have embraced a more holistic approach which joins these different facets to look at manuscripts as complex objects, created, used, read and modified over long periods of time.

Digital surrogates have introduced an unprecedented ease into viewing, comparing and analyzing manuscript pages. The recent Covid pandemic has accelerated the way we embed digital elements in manuscript research and teaching. As manuscript collections were closed and classes moved to virtual learning environments, teachers, students and researchers grew to rely on digital surrogates. Even tutors attached to well-endowed manuscript collections, who had had the fortune to employ physical manuscripts in their teaching, have begun embracing digital technologies. This has accelerated preexisting trajectories, although one cannot assume a uniform progress, as digitization of manuscripts has followed thematic and national trajectories, with many libraries still lacking a significant digital component. Online teaching has likewise built on past achievements, with palaeography modules such as the *album interactif de paléographie médiévale* and Public Record Office palaeography tutorial used by students both within and without the classroom.[2]

---

[1] I wish to thank the organisers and participants of the conference for allowing me to present parts of the following paper and receive important feedback. Key ideas were also presented at the workshops *The Digital Medieval Manuscript* (St Andrews 2021) and *Prague Medieval DH Storming* (2023). I am in debt to Lucie Doležalová and Suzanne Paul for reading and commenting on earlier drafts of this paper, and for Jan Odstrčilík and the anonymous reviewers for many useful comments.

A greater reliance on digital surrogates has often been accepted at face value, uncritically. This has been challenged recently. The launch of Parker 2.0, a web platform which introduced new features to the digitization of the manuscript collection at the Parker Library, Corpus Christi College, Cambridge, provided important reflections on the new technology and its impact. The ensuing volume offers critical reflections on the digitization of manuscripts and the use of digital surrogates.[3] It advocates not only a greater awareness to how 'the gaps and fissures between material object and print or digital realization open a productive space where we can think through how we relate to these objects'[4], but also the need of greater transparency in the process of digitization, a claim reiterated in a PhD dissertation by Suzette van Haaren.[5]

Digital images have become ubiquitous in manuscript studies. Their centrality has sidelined other innovations in the field. Some of the most exciting developments in the study of manuscripts come from scientific collaborations. The information retrieved by new technologies breaks new grounds in our analysis of the manuscript page. Whereas past technologies often confirmed and quantified what is evident to the naked eye or the trained scholar (e.g. the type of animal whose skin was used as parchment or the density of dirt left by readers and users), new technologies are also providing altogether unknown, and hitherto unknowable, information. They supply manuscript scholars with a range of data that would have otherwise been left completely invisible, be it the retrieval of hidden texts and images, the DNA profile of the animals whose skins were used as writing support, or unearthing later modification of the page. As technologies are becoming more widespread and affordable, they are gradually embraced by libraries and heritage institutions. Their proliferation is further enhanced by the development of non-invasive procedures (with a few, such as Proteomics and DNA analysis discussed below, being micro-invasive) which do not require physical contact or sampling. This simplifies use and eases gaining the curatorial approval necessary before deploying any new technology to a manuscript. Despite the hesitant acceptance of these new technologies by scholars and some research libraries, they irreversibly transform the way we conduct research, and we can no longer ignore, or consider them as optional addenda to the traditional field.[6]

\*\*\*\*\*\*\*\*\*\*\*\*\*\*\*\*\*

For more than a decade I have been collaborating with scientists, curators, imaging specialists and developers on new ways of researching manuscripts and early printed books. I have also been involved in a variety of teaching initiatives, inducting advanced students in working with manuscripts in a digital environment. This paper provides me with the opportunity to present some of these initiatives (mostly based on my experience in the UK and the EU), the journey they have entailed and their wider applicability. It also enables me to reflect about moving beyond such ad-hoc solutions to scalable and replicable models.

---

[2] https://paleographie.huma-num.fr; https://www.nationalarchives.gov.uk/palaeography/where_to_start.htm (accessed 05 May 2023).

[3] Benjamin Albritton, Georgia Henley, and Elaine Treharne, eds. *Medieval manuscripts in the digital age*, Digital research in the arts and humanities (London and New York:Routledge, 2021), and especially especially: Astrid J. Smith, 'What it is to be a Digitization Specialist: Chasing Medieval Materials in a Sea of Pixels', pp. 17-24; Siân Echard, 'Rolling with It: Navigating Absence in the Digital Realm', pp. 82-90; Catherine E. Karkov, 'Severed Heads and Sutured Skins', pp. 190-204.

[4] Echard, 'Rolling with It', p. 88.

[5] Suzette van Haaren, 'The Digital Medieval Manuscript: Approaches to Digital Codicology', PhD Dissertation, University of Groningen, 2022.

[6] An discussion of the variety of technologies available for scientific analysis of manuscripts took place in a workshop convened by Matthew Collins, Alexandra Franklin and Peter Stallybrass, Oxford, 2017, and summarized in https://www.science.org/content/article/goats-bookworms-monk-s-kiss-biologists-reveal-hidden-history-ancient-gospels (accessed 05 May 2023).

I was first introduced to the potential and mechanisms of collaborative technological research through the analysis of an early modern Bible.[7] Working on subsequent uses and transformations of the book, I was intrigued to discover that the margins and empty spaces of the book had been heavily annotated, and later pasted over by thick paper. Unfortunately, it proved nearly impossible to decipher these annotations. Invisible to the naked eye, they merge with the other side of the page when using a backlight. Six months of attempts with different types of image manipulation had ended in vain, given the similarity between the ink of both strata. Finally, Graham Davis, Professor of 3D X-ray Imaging in the School of Dentistry at Queen Mary University of London, provided the solution. Working together for a few days, we took fresh images with and without backlight and discussed possible solutions. This has led Graham to write an algorithm subtracting one image from the other, completely cleaning the image and fully retrieving the annotations (Figures 1-3 depict the three stages of analysis: normal light, backlight, and following digital subtraction). We were now able to read the annotations, which contain important information on the subsequent uses of the book, and, crucially, the hesitant course of religious reform in sixteenth-century England. Collaboration was key. We did not use any sophisticated or bespoke equipment. With only a standard DSLR and an inexpensive light sheet, commonly available in reading rooms, establishing cross-disciplinary communication was paramount. We needed to frame the problem in terms of image analysis, rather than historical context. We then worked together to solve the problem, jointly examining the book, liaising on subsequent photo shoots, and finally discussing methodology and analysis of results.

Problem-centred collaboration is only one possible model for joint scientific analysis. A wider range of technologies, and a more blue-sky approach, characterized a subsequent collaboration with Paola Ricciardi, then Senior Research Scientist at the Fitzwilliam Museum, Cambridge. Rather than coming up with a specific problem to be solved, I have identified historically significant objects with possible layers of transformations: Henry VIII's and Thomas Cromwell's presentation copies of the Great Bible (Paris/London 1538/9). We were able to secure one book, currently at St John's College, Cambridge, for a three-day study in the analytical labs at the Fitzwilliam Museum. During that time we applied a range of non-invasive technologies, while framing new research questions as the analysis progressed.

The title page of the book proved especially receptive to complex analysis. An important starting point has been a modification identified prior to the collaboration: an image of a woman which was transformed to resemble a prominent figure from Henry VIII's court, possibly Jane Seymour (Figure 4).[8] Raman spectroscopy and X-Ray Fluorescence (XRF) spectroscopy provide clear insight into the chemical compilation of compounds on the page.[9] This has enabled the identification of pigments, which joins more traditional art historical analysis in ascertaining artists and workshops working on the book (and, indirectly, assists in establishing date and provenance). Better appreciation of pigments, assisted also by near-IR imaging, enabled the reconstruction of the image as it would have been seen in the past, before subsequent modifications by generations of readers and the transformations of materials. This analysis has revealed, for example, that the woman's dress, which is currently drab grey, was originally dazzling silver. The oxidized silver has modified the appearance of the image nowadays, and its reconstruction changes the balance of the page, revealing the centrality of this person in the original composition (supported also by its use of gold leaf, rather than the shell gold employed elsewhere on the page).[10]

---

[7] Eyal Poleg, 'The First Bible Printed in England: A Little Known Witness from Late Henrician England', *Journal of Ecclesiastical History* 67:4 (October 2016), 760-80.

[8] See Eyal Poleg, *A Material History of the Bible, England 1200-1553* (Oxford:Oxford University Press, 2020), pp. 125-7

[9] An introduction to the range of technologies discussed here is: Paola Ricciardi and Catherine Schmidt Patterson, 'Science of the Book: Analytical Methods for the Study of Illuminated Manuscripts', *The Art & Science of Illuminated Manuscripts: A Handbook*, ed. Stella Panayotova (Turnhout:Brepols, 2021), pp. 35-72.

New technologies can further the analysis of elements initially identified by the naked eye. They can also provide information and reveal elements which cannot be detected in other means. The use of microscopes on the same page has revealed hitherto unknown elements within that celebrated Tudor Bible. Microscopy, often employed by conservators, also assists in identifying minute details in the style of scribes or artists. When combined with varying light sources it revealed modifications to the page (as could other, more complex, surface-mapping technologies). Figure 4, the title page of the presentation copy of Henry VIII's Great Bible, has commonly been used by scholars of the English Reformation. However, deploying a microscope with raking light (a light source at an oblique angle) has revealed that two of the faces had been painted on a separate piece of parchment and pasted onto the page at a later date (figure 5). This modification has been skillfully masked by the Tudor artist, and can only be revealed by the new technology. It adds a new dimension to our understanding of the manipulation of the page, and the court politics behind it, as the page was continuously transformed in dialogue with shifting Tudor politics and Henry VIII's attitude to lay Bible reading. These modifications, and others identified in the Bible, shed important light on its creation and use, about which little external information survives. They demonstrate the value of a more proactive collaborative model, which, once more, was facilitated by a fully joint working pattern.

Following these initial discoveries, and with the support of the Centre for Visual Cultures, Cambridge, we were able to transport the sibling-Bible from the National Library of Wales for analysis in the lab. This has unearthed some unique problems posed by this complex object, which, in turn, required new means of analysis. The book has pages which appear to imitate its original compilation, in which scribes and illuminators mimicked the printed page. This may be a later interpolation. However, pigment analysis suggests this might have been contemporaneous to the creation of the book.[11] Codicological analysis has identified specific bifolia (two connected pages), which were not part of the printed book. However, as they imitate the original campaign, assessing their relation to the original book proves challenging. Here, additional technologies may offer a clue. Proteomics (the study of protein to infer, in our case, on breeds or environment) and ancient DNA (aDNA) analysis are becoming some of the most powerful tools for codicological analysis.[12] Manuscripts are often written on parchment, processed skin which still retains information about the animal from which it was made. For bio-archaeologists the parchment pages of manuscripts and documents are vast depositories of biological information about ancient breeds. This has been spearheaded by the Beasts2Craft project (York/Cambridge/Copenhagen) and was aided by two recent developments. The exponential growth in the commercial and academic use of proteomics and DNA analysis has led to a significant reduction in the size of sample required, and a decrease in price, making this more and more available for scholars lacking the funding associated with research in medicine or the sciences. New sampling techniques introduced by Sarah Fiddyment employ the electrostatic qualities of a common rubber (eraser) to lift protein and DNA samples from a manuscript page.[13] This micro-invasive technology replaced the need to physically remove a piece of parchment, and eases approval by conservators and curators, although such approval is not granted by all libraries.

---

[10] For initial analysis see: Eyal Poleg and Paola Ricciardi, 'How Thomas Cromwell Used Cut and Paste to Insert Himself into Henry VIII's Great Bible', *The Conversation* 13 Aug. 2020, https://theconversation.com/how-thomas-cromwell-used-cut-and-paste-to-insert-himself-into-henry-viiis-great-bible-143765.

[11] We are currently preparing these results for publication in a forthcoming article.

[12] An example for the use of these technologies across a range of manuscript is Joanna Story's *The Insular Manuscripts* Project (https://le.ac.uk/insular-manuscripts/, accessed 27 April 2023).

[13] Sarah Fiddyment, Bruce Holsinger, Chiara Ruzzier, Alexander Devine, et al. 'Animal origin of 13th-century uterine vellum revealed using noninvasive peptide fingerprinting', *Proceedings of the National Academy of Sciences* 112:49 (2015), 15066-71.

The link between proteomics and in-depth codicological analysis has been practiced in a study of a privately-owned twelfth-century glossed Bible (whose owner's willingness to explore new technologies opened up new research opportunities). Unambiguous evidence for the different animals used for its parchment was linked with scribal practices and textual units to present a nuanced view of the book's creation and its codicology.[14] Proteomics has also been used to assess stains in manuscripts and rolls, inferring, for example, on possible uses of birthing scrolls. The technology is not without its limitations. Not every parchment supports full analysis, which could be hindered, for example, by its means of production. More data on ancient breeds and the parchment trade is needed to support the use of these technologies to date and localise manuscripts. In-depth discussion with Sarah Fiddyment (proteomics) and Matthew Teansdale (aDNA) have led us to adopt a more comparative approach, which has the potential to provide important insights. In the National Library of Wales's Bible, for example, we are currently employing aDNA analysis to question whether leaves were inserted at a later date or were part of the original compilation of a book. Swabbing adjacent leaves and looking whether these share a DNA profile would, hopefully, help in supporting this codicological inquiry in new ways. This is a prolonged process, and sequencing is currently taking place, with results forecasted in the coming months.

Reflecting upon these projects, informed and honest communication has been key in ensuring their success (and an important shortfall when recalling failed collaborations). It necessitated me to move away from seeing technical analysis as a service provided to scholars. Rather, it is a joint endeavour, done as a collaborative work – the historian needs to be present throughout the analysis, and engage in ongoing dialogue with scientists, curators and technicians. There is an important learning curve, and a need to understand key facets of analytical technologies, and especially their potential and limitations. The same is true for the scientists and technicians involved, who likewise benefit from knowledge of historical questions and methodologies. An appreciation of the limitations of scientific analyses has moved these projects away from ascertaining absolute information on date and provenance and into the wide possibilities of a more comparative approach, rewriting research questions and developing further methodologies.

**These collaborations demonstrate the potential of employing analytical technologies in the historical study of ancient books. While they have revealed hitherto unknown facets of their creation and use, I would not hurry to adopt a triumphalist tone. These were ad-hoc collaborations. With nearly zero-budget, they relied on the good will of colleagues, research teams, libraries and labs. They did not originate within institutional hubs, but rather in random meetings which grew to fuller collaborations. In many ways, they were not part of an institutional culture, but rather required us to break away from it, taking time off to sample, analyse and publish. They were made possible by being in privileged positions, on open-ended contracts and embedded in some of the world's foremost heritage institutions. Junior colleagues, those on precarious contracts, or outside leading research and heritage centres, would have found creating and running similar projects verging on the impossible.**

\*\*\*\*\*\*\*\*\*\*\*\*\*\*\*\*\*

Rapid advances in technological analyses of the page do not replace the manuscript scholar. New technologies cannot substitute traditional disciplinary training in palaeography, codicology, art history etc. Rather, they add new information to support further, often more complex and better informed, scholarly analysis. How do scholars, and the field at large, react to this challenge? A useful litmus test for assessing the place of new technologies within the field is the training offered to advanced students in manuscript studies. What is the place of the traditional skill set employed to study manuscripts, and how can we imagine working on manuscripts in the new reality? Such training (primarily at graduate and postgraduate levels) reveals not only the current state of the field, but also how we envision the next generation of scholars. It can also serve as a way of reimagining the future of the field, and its complexities.

---

[14] B. C. Barker-Benfield with Andrew Honey and William Zachs, *The Glossed Luke with the Letter A: A Manuscript from St Augustine's Abbey, Canterbury* (Edinburgh: Blackie House, 2020).

In many universities the separation between manuscript studies and Digital Humanities is manifested in teaching provisions. There, training in manuscript studies is commonly reserved to 'traditional' disciplines (e.g. palaeography, codicology, art history or philology), with any significant digital component being an add-on or optional modules. Digital modules likewise rarely include more traditional means of manuscript study, and often focus primarily on the digital perspective of manuscript analysis and presentation.

The need to integrate Digital Humanities and scientific analysis into manuscript studies is evident. Training the next generation of scholars and defining the field of manuscript studies requires another key consideration. Current training in manuscript studies is aimed at providing tools for active engagement with the manuscript. Palaeography, codicology, philology or art history supply advanced students with the practical skill-set necessary to actively study manuscripts. They enable students to critically reflect on the work of past scholars, while also, importantly, conducting independent research. The move to greater use of new technologies should follow a similar trajectory. Students should not be solely designated as recipients of these technologies. While courses do allude to the existence of new technologies, only a handful extend to train students in these technologies. We should support students in becoming both informed end-users of the new technologies, as well as design ways of integrating them into their own research. This should apply to the entire technological gamut, from working in imaging suites, through using commercial and custom-made tools, to the creation of new digital resources. Much like allowing students to specialize in palaeography or philology, they should also be able to develop their coding skills or employ the tools of heritage science. Unlike other skill-sets, however, they cannot be expected to manage this all on their own. Rather, by experimenting with these technologies in their own projects, students would develop a better and more informed appreciation of their potentials and limitations as active users in collaborative environments.

\*\*\*\*\*\*\*\*\*\*\*\*\*\*\*

The rise of new technologies leads to an inevitable conclusion. There is a greater need, if not a necessity, to reimagine manuscript studies as embedded within new forms of analysis. The traditional tools of manuscripts scholars — palaeography, codicology, stemmatology, or art-historical analysis — should be joined by coding, imaging technologies, spectroscopy and proteomics. This would support not only a more informed study of manuscripts in their fullest extent, but also the ability to develop and conduct truly interdisciplinary research projects.

Filling this lacuna is far from straightforward. Changes to graduate teaching and diminishing prospects of academic careers mean that many of our students would not be, nor should be solely taught to become, the academics of tomorrow. Shortening of MA programmes and working within a highly competitive market further complicates the place of manuscript studies. Even without this additional plethora of training requirements, students in intensive manuscript courses experience a training-overload. Unlike some other degree programmes in the humanities (common in the UK and the US), in which training requirements are minimal, students undertaking a manuscript-centred degree are often required to master ancient and modern languages, as well as manuscript-specific skill-sets, from deciphering historical scripts to reconstructing the compilation of a codex. Such training is complex and cannot be taught en masse. In an academic environment where manuscript-centred degrees are already in a precarious position, increasing the demand on students could make them completely untenable. Furthermore, fully embracing new technologies would move students, who have been training within the humanities, out of their comfort zone. It would require them to adjust to information technologies and the natural sciences, with their often-alien methods and discourses.

I have been running digitally oriented training programmes for advanced students in manuscript studies for more than a decade. These have provided me with an opportunity to develop and reflect on methodological needs, potential and limitations, while receiving important feedback from generations of students. Some of the programmes were embedded into universities' curriculum, while others were multi-national training opportunities; some catered for a specific skill-set while others were more reactive and versatile. As a whole, they gradually moved beyond traditional teaching provision to begin reimagining our training offering. Most of our modules are still taught in weekly sessions. There, teachers aim to provide an entire class with a mostly uniform skill-set necessary for consulting and researching manuscripts. Experimenting with a range of teaching programmes has led me to see the possibility, if not the necessity, of moving beyond this paradigm to embed the use of new technologies within our curriculum. This leads us away from thinking of manuscript studies, Digital Humanities and new technologies as separate disciplines.

My first brush with teaching new technologies in manuscript studies was an optional module offered to MSt students in Oxford (2012/13). *Digital Manuscripts* introduced graduate students to the theory and practice of digital editions in the first term. In the second, Oxford's Teaching Project Award funded a developer to work together with the students on their editions. This, I regret to say, has never materialised. Despite some students preparing fascinating editions, the work with the developer stalled and then retracted altogether. Change of personnel, the two-stage plan, and the lack of allocated time to liaise and supervise this work, have resulted in failure. Together with other complications arising from working alongside developers in past projects, this had demonstrated to me that collaborations with developers cannot be taken for granted. Such work should not be seen as an add-on component, additional to the core of teaching programmes or research projects. Rather, there is a clear need for students (and faculty) to be gradually initiated into collaboration with IT specialists in a supportive environment, a key part of training in the advanced technological study of manuscripts.

This collaborative approach was developed in a three-year training programme, generously funded by the EU's Erasmus+ scheme: *Digital Editing of Medieval Manuscripts* (*DEMM*, 2014-17). Each year, students from participating universities across Europe came together for three one-week training schools, each week introducing an area less commonly studied in manuscript classes. The first week provided training in philology, initiating students in critical editions and supporting their own research on medieval texts. The second week was an intensive immersion in the Textual Encoding Imitative (TEI) which provides templates for encoding scholarly editions in XML. Students then went back to their home institutions and were supported in encoding part of an edition they were working on. In the last week the students brought their encoded text and liaised with developers to prepare it as an online edition. The programme created a library of short scholarly editions (subsequently published on https://www.digitalmanuscripts.eu), while providing training opportunities and CV-enhancement for students, many of whom continued in the field of manuscript studies and digital technologies.

The rationale of the training programme – short bursts of teaching and workshops interspersed with independent work of students – started as a practical necessity. It was the only way of bringing students from across Europe to study together with experts in discrete fields. Over time, however, it proved to be a meaningful learning experience, not inferior to that achieved in a more traditional classroom. Students were able to ingest considerable amount of information, experiment innovatively with new concepts, and liaise with faculty and peers in meaningful ways. They also benefitted from cross-national networking opportunities.

Running the programme was complex, not the least due to its nebulous admin and finance (whose brunt was selflessly bore by Lucie Doležalová, Charles University Prague). However, it was the third week which proved most challenging. The other two weeks were modeled on more traditional training programmes. The third, however, was a novelty to us. Initial attempts to work alongside professional developers or advanced Computer Science students proved too challenging for all those involved. Communication problems and managing expectations became paramount. To assist in achieving a meaningful collaboration within a short span of time, we resorted to working alongside developers with some background in the humanities, which significantly eased communications with students. The funding of the project uniquely enabled students to travel across Europe, providing accommodations, and – importantly – paid for developers and admin support. It also added a significant burden for the organisers, and limited the scope of the programme to three years, preventing fully embedding it in each university's curriculum.

When funding ran out, the convenors wished to continue providing similar training, leading to two separate training projects: *Digital Scholarly Editions: Manuscripts, Texts, and TEI Encoding*, a week-long school for students from Lyon, London, Prague, Budapest, Bratislava and Tokyo, was convened by Marjorie Burghart and ran as a 'flipped classroom' – online course followed by in-site practical sessions. Yearly one-week immersion classes on TEI were funded by the French Regional Government, and brought students and staff together in a combination of practical workshops and project presentations. Its funding ended in 2021.

*Hands:on* is an annual teaching collaboration between Queen Mary University of London and Cambridge University Library.[15] Students from the UK, Europe and beyond join in a week-long hackathon, working with curators and developers to produce digital prototypes for consulting manuscripts online. Each year's hackathon has a different theme, linked to current or forthcoming exhibitions or research projects (which we term 'client'). Thus, for example, the 2019 hackathon supported the Palimpsest exhibition at Cambridge University Library, the 2022 hackathon (following a two-year Covid hiatus) linked to a forthcoming Silk Roads exhibition at the British Museum, and in 2023 we worked alongside the *Curious Cures* project on medieval medical recipes. The programme does not aim to produce fully functional products, but rather proofs of concept – innovative prototypes which present novel means of engaging with manuscripts online.

Students receive light-touch online training in soft-skills such as product design and project management prior to the hackathon. They then divide into groups, ensuring each group has members with and without coding experience, outreach or manuscript-related skills. Each group chooses a user persona (such as a retired schoolteacher, a bored teenager, or an academic researcher) and creates a prototype for them. The groups first create a mockup and then digital prototypes, which are user-tested before finalised and presented to a wide audience of academics and non-HEI staff. The 'client' is then supported in utilising the prototypes and transforming them into fully functional web resources. The funding structure of the programme breaks away from past initiatives, as it does not rely on external funding, nor on students paying their own expenses. Each participating university funds its own students, and additional costs are kept at a minimum (and mostly covered by the project's 'client'). This has enabled us to avoid short-term funding cycles. As the programme is cost-neutral, we can keep it open-ended, enhancing the possibility of incorporating it into universities' curriculum (which is still a desideratum).

*****************

---

[15] https://projects.history.qmul.ac.uk/handson/, accessed 17 May 2023. The programme is convened by Mary Chester-Kadwell and Suzanne Paul (Cambridge University Library), Chris Sparks and Eyal Poleg (Queen Mary University of London).

Some of the abovementioned collaborations started in running clubs or playground talks; they were ad-hoc initiatives based on past acquaintanceship or luck. How to transition these disparate research and training initiatives into scalable and replicable programmes? Training is key – teaching how to efficiently collaborate with scientists, developers and other non-academics equips the next generation of scholars with important tools, while introducing them to current research and dissemination initiatives. Universities celebrate research-led teaching. Here is an opportunity to engage in teaching-led research, in which collaborations created in teaching programmes support research of advanced students and faculty alike. The combined experience of these training programmes has led me to identify key traits for rethinking how to integrate innovative training within manuscript studies:

1. **Modular**. To avoid a training overload, there is a need to reimagine our teaching provision. Rather than providing a blanket training for entire classes, training should be agile, on-demand, and, inevitably, partial. Most students, for example, do not need to learn the entire gamut of Western scripts 500-1500, nor full knowledge of XRF spectra analysis. Both traditional manuscript skills and implementation of new technologies should be tailored to students' projects, and would, when brought together, provide students with a unique and bespoke learning experience. Being responsive to students' educational needs would lead to a tuition that is more relevant, and better suited, for emerging projects and career development. Such training would initiate students in conducting interdisciplinary research projects, which would better serve them both within their degree programmes and hereafter.
2. **Flexible**. Some topics, especially those requiring repetition and long gestation periods, are perfectly suited for weekly classes. Others, however, are not. Past training programmes have surprised us in showing that intensive weeks provide an equally significant learning experience to weekly seminars and lectures, as does the combination of online and on-site training. This type of immersion experience assists in initiating students to new types of work. This is hardly a novelty, as such models have been used by students and practitioners in computing and the sciences, where the model of the hackathon was developed. Such flexibility also enables bringing together a wide range of expertise for a limited period of time. Bringing together developers, curators, conservators and advanced students is more likely to occur in a condensed week training rather than in term-long weekly sessions.
3. **Cross-Institutional.** These are complex, bespoke and often niche training programmes. One would often struggle to recruit enough students from a single institution to justify the investment of staff time and additional resources necessary for an immersive training programme. National and international collaborations ensure full attendance by keen and suitable students while supporting on-demand and bespoke learning. They require an additional level of administration and support, while going against some universities' managerial imperatives and financial structures. Important side-effects are networking opportunities and establishing future scholarly communities and collaborations. Only by stopping to think about each university as an isolated bastion of learning can we truly embed the multitude of disciplines necessary for manuscript studies in the new era.
4. **Practical**. Embracing the Hackathon model leads us to rethink goals and assignments. Universities often employ imagined assignments, in which students' work is viewed only by, and relevant only to, the module's tutor. Some manuscript modules have moved away from this, harnessing their students' work to further knowledge more widely, as in providing support for manuscript cataloguing. New technologies lend themselves more easily to practical implementations, which is a model we have embraced in different training programmes. This could either be students' individual projects, which they then publish online to further academic knowledge, provide them with early scholarly engagement, and serve as a significant boost for their CVs. It could also lead to harnessing an entire training programme to support a research project or an exhibition, in which students gain practical experience and a significant output. The commissioning institution is offered a unique way of improving their digital environment while benefitting from the necessary time investment to engage with key aspects, such as product design or user testing, which are often sidelined in academic projects.
5. **Collaborative**. Moving to a truly cross-disciplinary training programme requires a wide array of skills. These are beyond the remit of a single individual, and are often not found even within the same department or institution. Training should be delivered by, and with, historians, palaeographers, philologists, scientists, curators and developers. It should bring together universities, libraries, archives and museums. Past experience has revealed this to be a challenging endeavour, and one which requires careful planning and execution. One cannot take efficient cross-disciplinary work and collaboration for granted; it must be managed and maintained. Such a collaborative experience, however, has an immense value in of itself and serves as an important learning opportunity for students, preparing them for work both within and without academia.
6. **Soft Skills**. This topic is often sidelined within university training, which tends to prioritise hard disciplinary skills such as palaeography, codicology, or languages. However, reflecting on past training programmes and their long-term benefit for students, it is evident that soft skills take a central role and serve students' future endeavours. Project- and time-management, communication skills and product design are all crucial for collaborative work. They are fully transferable, and invaluable for managing research projects. Less explicitly supported in humanities studies, they cannot be taken for granted. In past training programmes we taught these through a combination of online modules and on-site hands-on training, which provided an opportunity to experiment in a supportive and reflexive environment.

\*\*\*\*\*\*\*\*\*\*\*\*\*\*\*

The growing reliance on digital surrogates and data analysis tools has often been accepted uncritically in research and teaching. Their use has also advanced the field in specific trajectories. Manuscripts – ancient, medieval and modern – are not only depositories of texts. Viewing manuscripts as complex objects quickly moves us beyond the primarily textual and image-based tools used by scholars of Digital Humanities. It leads us to deploy technologies which are often the reserve of archaeologists and conservators. These are practiced by imaging specialists and heritage scientists, who tend to be affiliated with museums and conservation departments. The potential they hold for manuscript studies is, nevertheless, immense. As new technologies proliferate, the existence of manuscript studies as removed from Digital Humanities and heritage science is becoming more and more artificial and detrimental to the future of the field.

While often seen as a stronghold of traditional teaching, manuscript studies has been one of the most innovative and forward-looking areas of academic pursuit from its onset. Embracing new teaching initiatives and integrating new technologies into manuscript studies is now a necessity. It also has the potential of making the field a model for humanities teaching in the twenty-first century. This transformation is complex and necessitates reimagining the field, the way we research and teach. It needs to start with the teachers. The EU's educational strategies of *Teaching the Future* and *Pedagogics to the Max* highlight the high expectations this sets on educators, moving them beyond their comfort zone and into new pedagogical realities.[16] Any transformation of the field would need to be in dialogue with teaching staff, bringing their expertise and practices into its remit. This transformation will provide an immense service to our students, their academic capacities and future employability. It will also serve to change the way we study, teach and research, enabling us to truly embrace the future – and present – of manuscript studies.

---

[16] https://knowledge4policy.ec.europa.eu/diversification-education-learning_en#megatrend, accessed 04 May 2023.